\documentclass{svjour3}                     
\smartqed  
\usepackage{graphicx}
\begin{document}

\title{The SMBH mass versus $M_{\mathrm{G}} \sigma^2$ relation: A comparison between real data and numerical models}

\titlerunning{The $M_{\bullet}$ versus $M_{\mathrm{G}} \sigma^2$ relation}        

\author{Antonio Feoli         \and
        Luigi Mancini         \and \\
        Federico Marulli      \and
        Sidney van den Bergh
}


\institute{A. Feoli \at
              Dipartimento di Ingegneria, Universit\`{a} del Sannio, \\
              Corso Garibaldi n. 107, Palazzo Bosco Lucarelli, 82100–-Benevento, Italy \\
              \email{feoli@unisannio.it}           
           \and
           L. Mancini \at
              Dipartimento di Fisica ``E.R. Caianiello'', Universit\`{a} di Salerno, \\
              Via Ponte Don Melillo, 84084–-Fisciano (SA), Italy \\
              Istituto Nazionale di Fisica Nucleare, Sezione di Napoli, Italy \\
             Istituto Internazionale per gli Alti Studi Scientifici (IIASS), Vietri Sul Mare (SA), Italy \\
              \email{lmancini@physics.unisa.it}           
           \and
           F. Marulli \at
              Dipartimento di Astronomia, Universit\`{a} di Bologna, \\
              Via Ranzani 1, 40127--Bologna, Italy \\
              \email{federico.marulli3@unibo.it}           
             \and
           S. van den Bergh \at
              Dominion Astrophysical Observatory, Herzberg Institute of Astrophysics, \\
              5071 West Saanich Road, Victoria, BC, V9E 2E7, Canada \\
              \email{sidney.vandenbergh@nrc-cnrc.gc.ca}           
}

\date{Received: date / Accepted: date}
\maketitle

\begin{abstract}
The relation between the mass of supermassive black holes located in the center of the host galaxies and the kinetic energy of random motions of the corresponding bulges can be reinterpreted as an age--temperature diagram for galaxies. This relation fits the experimental data better than the $M_{\bullet}-M_{\mathrm{G}}$, $M_{\bullet}-L_{\mathrm{G}}$, and $M_{\bullet}-\sigma$ laws. The validity of this statement has been confirmed by using three samples extracted from different catalogues of galaxies. In the framework of the $\Lambda$CDM cosmology our relation has been compared with the predictions of two galaxy formation models based on the Millennium Simulation.


%
\end{abstract}
%

\section{Introduction}
\label{sec:1}
Nowadays an increasingly number of multi-band observational evidences reinforces the picture that all massive galaxies, with a spheroidal component, contain a super massive black hole (SMBH, $M_{\bullet} > 10^6 \, M_{\odot}$) at their center. Some kind of connection between the traits of central SMBHs and the evolutionary state of their host galaxies should be expected and actually it is currently under study and debate. In particular, the mass of the SMBHs seems to be closely related with the properties of the spheroids in which they reside. Many efforts have been made in this sense in order to achieve a pragmatic correlation, such as for example the ones between the mass of the SMBH and the bulge\footnote{Here \emph{bulge} refers to either the spheroidal component of a spiral/lenticular galaxy or to a full elliptical galaxy.} stellar mass or luminosity ($M_{\bullet}-M_{\mathrm{G}}$; $M_{\bullet}-L_{\mathrm{G}}$)~\cite{kormendy95,richstone98,magorrian98,laor01,wandel02,marconi03,haring04}, velocity dispersion ($M_{\bullet}-\sigma$)~\cite{ferrarese00,gebhardt00,tremaine02}, effective radius ($M_{\bullet}-R_{\mathrm{e}}$)~\cite{marconi03}, kinetic energy ($M_{\bullet}-M_{\mathrm{G}}\sigma^2$)~\cite{feoli05,feoli07}, S\'{e}rsic index ($M_{\bullet}-n$)~\cite{graham05,graham07}. However, as already remarked by Novak et al.~\cite{novak06}, it is difficult to understand the fundamental nature of the correlations between SMBH and host properties, because all such relations depend critically on the accuracy of the published error estimates in all quantities under consideration. Moreover, the conservative practice of adopting large error bars turned out not to be such a good strategy if only few data are considered for the fit.

A more complex approach is to read the above-mentioned relations in terms of a fundamental plane, which relates the SMBH mass to two or more spheroid properties such as galaxy effective radius, stellar velocity dispersion, luminosity, and mass~\cite{gebhardt00,hopkins07a}. A hydrodynamical simulation of major galaxy mergers, including the effects of black hole (BH) accretion and feedback, supports this picture~\cite{hopkins07b}. Other authors reached the same conclusion by modeling the cosmological co-evolution of galaxies and their central SMBHs within a semianalytical framework~\cite{marulli08}.


The possibility that the mass of a SMBH correlates with the total gravitational mass of its host galaxy, or with the mass of the dark matter halo in which it presumably formed, has been investigated~\cite{silk98,ferrarese02}, and supported by recent self-consistent simulations of the co-evolution of the SMBH and galaxy populations~\cite{booth09}, which asserts that the mass of a SMBH is determined primarily by the mass of the dark matter halo.

Also important is to understand the mechanisms that regulate the growth rate of the SMBH. Studies of the distribution of Eddington luminosity ratios, $L_{\mathrm{bol}}/L_{\mathrm{Edd}}$, of active galactic nuclei (AGNs) show that the energy-storing rate in luminous AGNs are ultimately determined by SMBH self-regulation of the accretion flow and this process acts on large scales ($>1$ kpc)~\cite{kollmeier06}.

An alternative fundamental plane has been proposed in order to correlate the radio and the X-ray luminosity, and the BH mass~\cite{merloni03,falcke04}. However, this plane turns into an effective BH mass predictor, like the very popular $M_{\bullet}-\sigma$ relation, only if the obscured AGNs are excluded~\cite{gultekin09b}. Actually, it is not clear at all if the intrinsic activity (Seyfert, Liner, etc.) of galaxies gives them a special place in the fundamental planes as well as in any diagram originated by the other proposed relations. This topic definitely deserves a deep investigation.

Very recently, Feoli \& Mancini~\cite{feoli09} (thereafter FM) suggested a fascinating connection between the stored energy of central SMBHs and the evolutionary process of galaxies. Using a sample of 64 galaxies, they investigated the relation between the mass of the SMBHs and the kinetic energy of the random motion of the corresponding galaxy bulges. Besides the fact that, as already noted in previous papers~\cite{feoli05,feoli07}, this relation works better than the most common $M_{\bullet}-\sigma$ and $M_{\bullet} - M_{\mathrm{G}}$ laws, they found some analogies between the $M_{\bullet}-M_{\mathrm{G}}\sigma^2/c^2$ plane for galaxies and the Hertzsprung--Russell (HR) diagram for stars. The HR diagram connects the energy radiated by the nucleus of a star with its surface temperature. Similarly, the FM diagram essentially connects a property of the inner nucleus of a galaxy, the energy stored by the SMBH, $M_{\bullet}c^2$, with a property of the external surface of its bulge, i.e., the kinetic energy of random motions. Moreover, each morphological type of galaxy generally occupies a different area in the FM plane, reminiscent of the different positions occupied by stars of the various spectral classes in the HR diagram.

In this paper, we want to verify the goodness of the $M_{\bullet}-M_{\mathrm{G}}\sigma^2$ relation, first of all comparing it with the other popular relations (essentially $M_{\bullet}-M_{\mathrm{G}}$, $M_{\bullet}-L_{\mathrm{G}}$, and  $M_{\bullet}-\sigma$). In order to do this, we consider three samples of galaxies extracted from three different catalogues. Thanks to this procedure, we avoid the possibility that the tightness of the FM relation might depend on a suitable choice of the data. Then, we complete our analysis comparing the results, derived by real data, with the predictions of two semi-analytic hierarchical models that follow the cosmological co-evolution of dark matter, galaxies and black holes.

The paper is structured as follows: in \S~2 we provide our reference catalogues and report the data of the correspondent samples.
In \S~3 we discuss the results which emerge from our analysis. These results are then compared with the predictions of two hierarchical models of galaxy structure formation in \S~4. Finally, \S~5 contains the summary of the main conclusions of this work.

\section{The samples}
\label{sec:2}
In order to check which relation is the most effective to predict the black hole mass, we have tested the $M_{\bullet}-\sigma$, the $M_{\bullet}-M_{\mathrm{G}}$, the $M_{\bullet}-L_{\mathrm{G}}$ and the $M_{\bullet}-M_{\mathrm{G}}\sigma^2$ relations on three different samples of galaxies extracted from the catalogues of Graham~\cite{graham08}, G\"{u}ltekin et al.~\cite{gultekin09}, and Hu~\cite{hu09} respectively. Since these three catalogues have been compiled independently, we are quite safe from any tampering of the data due to a personal intervention, that amounted to selection from the literature of the best values for the mass of the galaxies, which are missing in the first two catalogues.

The relations that we want to study can be written in the following form

\begin{equation}
\log_{10}{(M_{\bullet})} = b + m \log_{10}{x},
\label{Eq:01}
\end{equation}
where $m$ is the slope, $b$ is the normalization, and $x$ is a parameter of the host bulge such as the mass ($M_{\mathrm{G}}$), the luminosity ($L_{\mathrm{G}}$), the central velocity dispersion ($\sigma_{\mathrm{c}}$), the effective velocity dispersion ($\sigma_{\mathrm{e}}$), or the kinetic energy of the random motions ($M_{\mathrm{G}}\sigma^2/c^2$). Eq.~\ref{Eq:01} can be used to predict the values of $M_{\bullet}$ in other galaxies once we know the value of $x$. In order to minimize the scatter in the quantity to be predicted, we have to perform an ordinary least-squares regression of $M_{\bullet}$ on $x$ for the considered galaxies of which we already know both the quantities. In Table~\ref{tab:1} we collect the fits obtained for  all the relations for each sample. As in~\cite{feoli07,graham07,feoli09}, these fits were obtained taking into account the error bars in both variables and using the routine FITEXY~\cite{press92} for the relation $y = b + mx$, by minimizing the $\chi^2$.

\begin{table}
\caption{Black hole -- bulge correlations and fitting parameters}
\label{tab:1}       
\begin{tabular}{lcccrccc}
\hline\noalign{\smallskip}
Sample & \begin{tabular}{c} Number of \\ galaxies \\ \end{tabular} & Relation & $m\pm\mathrm{\Delta} m$ & $b\pm\mathrm{\Delta} b \; \; \;$ & $\chi_{\mathrm{r}}^2$ & $\varepsilon_{0}$& $r$\\
\noalign{\smallskip}\hline\noalign{\smallskip}
Graham 2008 & 59 & $M_{\bullet}-\sigma$ & $5.26\pm0.13$ & $8.20\pm0.02$ & 6.09 & 0.39 & 0.86 \\
Graham 2008 & 59 & $M_{\bullet}-M_{\mathrm{G}}$ & $1.18\pm0.05$ & $-4.77\pm0.51$ & 1.81 & 0.23 & 0.92 \\
Graham 2008 & 59 & $M_{\bullet}-M_{\mathrm{G}}\sigma^2$ & $0.83\pm0.04$ & $4.37\pm0.16$ & 1.18 & 0.12 & 0.94 \\

G\"{u}ltekin et al. 2009 & 55 & $M_{\bullet}-\sigma$ & $4.99\pm0.14$ & $8.26\pm0.03$ & 8.85  & 0.47 & 0.82 \\
G\"{u}ltekin et al. 2009 & 55 & $M_{\bullet}-M_{\mathrm{G}}$ & $1.22\pm0.05$ & $-5.09\pm0.57$ & 2.48 & 0.33 &  0.88 \\
G\"{u}ltekin et al. 2009 & 52 & $M_{\bullet}-L_{\mathrm{V}}$ & $1.48\pm0.08$ & $-7.16\pm0.82$ & 3.20 & 0.59 & 0.69 \\
G\"{u}ltekin et al. 2009 & 55 & $M_{\bullet}-M_{\mathrm{G}}\sigma^2$ & $0.86\pm0.04$ & $4.31\pm0.19$ & 1.68 & 0.24 & 0.90 \\

Hu 2009 & 58 & $M_{\bullet}-\sigma$ & $5.83\pm0.15$ & $8.20\pm0.02$ & 6.22 & 0.40 & 0.87 \\
Hu 2009 & 58 & $M_{\bullet}-M_{\mathrm{G}}$ & $1.27\pm0.04$ & $-5.57\pm0.49$ & 5.07 & 0.46 & 0.83 \\
Hu 2009 & 58 & $M_{\bullet}-L_{\mathrm{K}}$ & $1.48\pm0.03$ & $-7.88\pm0.29$ & 35.54 & 0.60 & 0.75 \\
Hu 2009 & 58 & $M_{\bullet}-M_{\mathrm{G}}\sigma^2$ & $0.91\pm0.04$ & $4.16\pm0.20$ & 2.19 & 0.34 & 0.86 \\

\noalign{\smallskip}\hline
\end{tabular}
\end{table}

\subsection{Graham sample}
\label{sec:2.1}
Graham (2008) compiled a catalogue of 76 galaxies with direct SMBH mass measurements~\cite{graham08}. We select only 61 galaxies (27 ellipticals, 19 lenticulars, 15 spirals), since a reliable value for the central velocity dispersion and the mass was not available for all the host bulges of the catalogue. These galaxies are listed in Table~\ref{tab:2} together with the values of their respective parameters. Concerning the errors in the measures, we adopt the same strategy as in~\cite{feoli05,feoli07}: We consider that the error for the bulge mass is 0.18 dex in $\mathrm{log}_{10}{M_{\mathrm{G}}}$ for all the galaxies~\cite{haring04}, while the relative error on the velocity dispersions is $5\%$.

\begin{figure}
\includegraphics[width=0.8\textwidth]{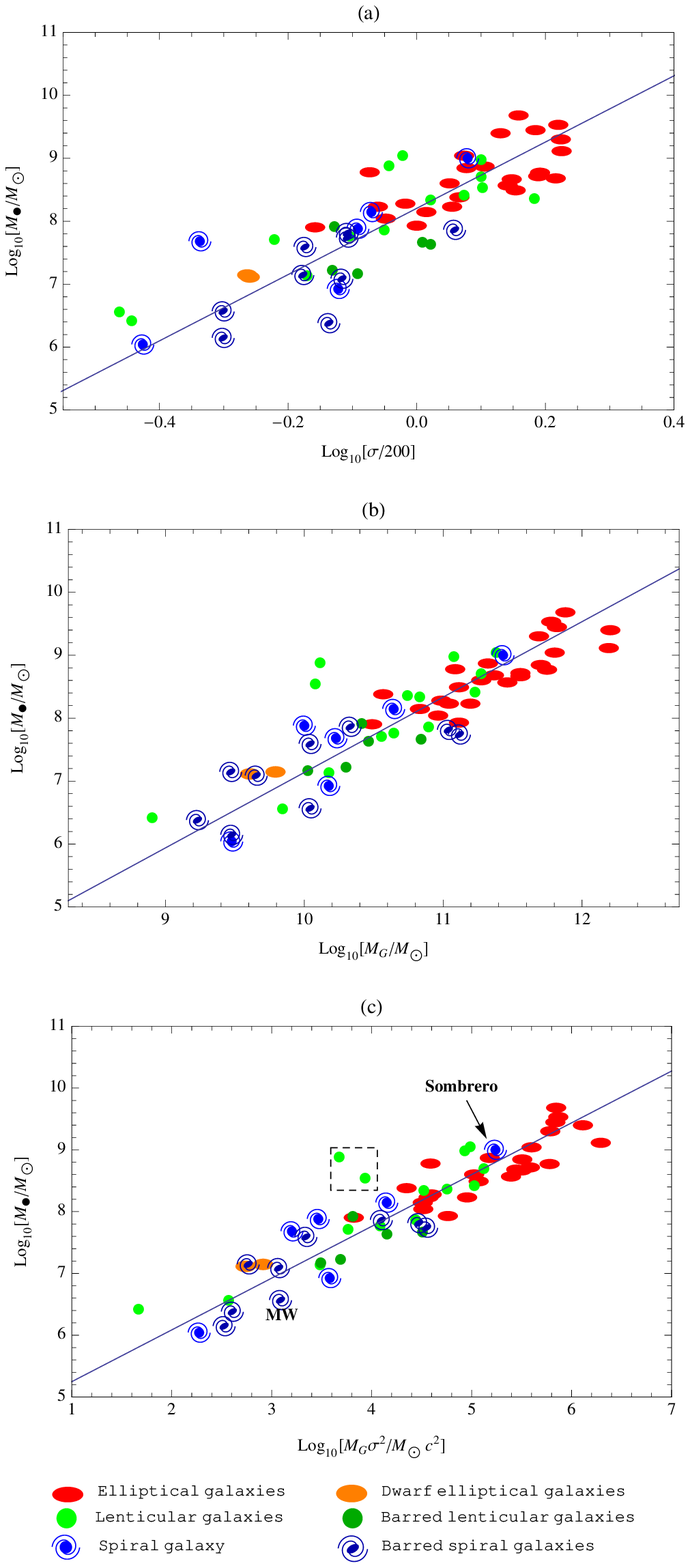}
\caption{Best-fitting (a) $M_{\bullet}-\sigma$, (b) $M_{\bullet}-M_{\mathrm{G}}$, and (c) $M_{\bullet}-M_{\mathrm{G}}\sigma^2$ relations for
the sample extracted from Graham (2008) \cite{graham08}. The symbols represent elliptical galaxies (red ellipses), lenticular galaxies (green circles), barred lenticular galaxies (dark green circles), spiral galaxy (blue spirals), barred spiral galaxies (dark blue barred spirals), and dwarf elliptical galaxies (orange ellipses). The dashed box encloses the lenticular galaxies NGC4342 and NGC4350, which have been excluded from the fits due to their high value of the $M_{\bullet}/M_{\mathrm{G}}$ ratio.}
\label{fig:1}       
\end{figure}

\begin{table}
\caption{The data are taken from Graham (2008)~\cite{graham08}, except for the values of $M_{\mathrm{G}}$ whose references are given in the last column. The value of $\sigma_{c}$ for Abell 1836 is taken from~\cite{gultekin09}. The error for the bulge mass is 0.18 dex in the $\log_{10} {M_{\mathrm{G}}}$ for all galaxies, while the relative error on the velocity dispersions is $5\%$. Brightest cluster galaxy (BCG) is defined as the brightest galaxy in a cluster of galaxies}
\label{tab:2}       
\begin{tabular}{lcccccc}
\hline\noalign{\smallskip}
Galaxy & Type & $\sigma_{\mathrm{c}} (km/s)$ & $M_{\bullet} (M_{\odot})$ & $\delta M_{\bullet} (M_{\odot})$ & $M_{\mathrm{G}} (M_{\odot})$ & References\\
\noalign{\smallskip}\hline\noalign{\smallskip}
Abell 1836 & BCG & 288 & $4.8 \times 10^9$ & $0.8 \times 10^9$ & $7.9 \times 10^{11}$ & \cite{hu09} \\
A3565/IC 4296 & BCG & 336 & $1.3 \times 10^9$ & $0.4 \times 10^9$ & $1.6 \times 10^{12}$ & \cite{dallabonta07}\\
CygnusA & E  & 270 & $2.5 \times 10^9$ & $0.7 \times 10^9$ & $1.6 \times 10^{12}$ & \cite{marconi03} \\
IC1459  & E3 & 306 & $2.8 \times 10^9$ & $1.2 \times 10^9$ & $6.6 \times 10^{11}$ & \cite{marconi03} \\
NGC821  & E  & 200 & $8.5 \times 10^7$ & $3.5 \times 10^7$ & $1.3 \times 10^{11}$ & \cite{haring04} \\
NGC1399 & E  & 329 & $4.8 \times 10^8$ & $0.7 \times 10^8$ & $2.3 \times 10^{11}$ & \cite{houghton06} \\
NGC2974 & E  & 227 & $1.7 \times 10^8$ & $0.3 \times 10^8$ & $1.6 \times 10^{11}$ & \cite{cappellari06} \\
NGC3377 & E5 & 139 & $8.0 \times 10^7$ & $0.6 \times 10^7$ & $3.1 \times 10^{10}$ & \cite{cappellari06} \\
NGC3379 & E  & 207 & $1.4 \times 10^8$ & $2.7 \times 10^8$ & $6.8 \times 10^{10}$ & \cite{haring04} \\
NGC3608 & E2 & 192 & $1.9 \times 10^8$ & $1.0 \times 10^8$ & $9.7 \times 10^{10}$ & \cite{haring04} \\
NGC4261 & E2 & 309 & $5.2 \times 10^8$ & $1.1 \times 10^8$ & $3.6 \times 10^{11}$ & \cite{haring04} \\
NGC4291 & E2 & 285 & $3.1 \times 10^8$ & $2.3 \times 10^8$ & $1.3 \times 10^{11}$ & \cite{haring04} \\
NGC4374 & E  & 281 & $4.6 \times 10^8$ & $3.5 \times 10^8$ & $3.6 \times 10^{11}$ & \cite{haring04} \\
NGC4473 & E5 & 179 & $1.1 \times 10^8$ & $0.8 \times 10^8$ & $9.2 \times 10^{10}$ & \cite{haring04} \\
NGC4742 & E4 & 109 & $1.4 \times 10^7$ & $0.5 \times 10^7$ & $6.2 \times 10^9$ & \cite{haring04} \\
NGC4486 & E0 & 332 & $3.4 \times 10^9$ & $1.0 \times 10^9$ & $6.0 \times 10^{11}$ & \cite{haring04} \\
NGC4486a & E & 110 & $1.3 \times 10^7$ & $0.8 \times 10^7$ & $4.1 \times 10^9$ & \cite{nowak07}  \\
NGC4486B & E & 169 & $6.0 \times 10^8$ & $3.0 \times 10^8$ & $1.2 \times 10^{11}$ & \cite{nowak07}  \\
NGC4621 & E  & 225 & $4.0 \times 10^8$ & $0.6 \times 10^8$ & $1.9 \times 10^{11}$ & \cite{cappellari06} \\
NGC4649 & E1 & 335 & $2.0 \times 10^9$ & $0.6 \times 10^9$ & $4.9 \times 10^{11}$ & \cite{haring04} \\
NGC4697 & E4 & 174 & $1.7 \times 10^8$ & $0.2 \times 10^8$ & $1.1 \times 10^{11}$ & \cite{haring04} \\
NGC5077 & E  & 255 & $7.4 \times 10^8$ & $4.7 \times 10^8$ & $2.1 \times 10^{11}$ & \cite{defrancesco08} \\
NGC5813 & E  & 239 & $7.0 \times 10^8$ & $1.1 \times 10^8$ & $5.1 \times 10^{11}$ & \cite{cappellari06} \\
NGC5845 & E3 & 233 & $2.4 \times 10^8$ & $1.4 \times 10^8$ & $3.7 \times 10^{10}$ & \cite{haring04} \\
NGC5846 & E  & 237 & $1.1 \times 10^9$ & $0.2 \times 10^9$ & $6.4 \times 10^{11}$ & \cite{cappellari06} \\
NGC6251 & E  & 311 & $5.9 \times 10^8$ & $2.0 \times 10^8$ & $5.6 \times 10^{11}$ & \cite{haring04} \\
NGC7052 & E4 & 277 & $3.7 \times 10^8$ & $2.6 \times 10^8$ & $2.9 \times 10^{11}$ & \cite{haring04} \\
\noalign{\smallskip}\hline\noalign{\smallskip}
NGC221  & S0 & 72  & $2.5 \times 10^6$ & $0.5 \times 10^6$ & $8.0 \times 10^8$  & \cite{haring04} \\
NGC3115 & S0 & 252 & $9.1 \times 10^8$ & $1.0 \times 10^9$ & $1.2 \times 10^{11}$ & \cite{haring04} \\
NGC3245 & S0 & 210 & $2.1 \times 10^8$ & $0.5 \times 10^8$ & $6.8 \times 10^{10}$ & \cite{haring04} \\
NGC3414 & S0 & 237 & $2.5 \times 10^8$ & $0.4 \times 10^8$ & $1.7 \times 10^{11}$ & \cite{cappellari06} \\
NGC3998 & S0 & 305 & $2.2 \times 10^8$ & $2.0 \times 10^8$ & $5.5 \times 10^{10}$ & \cite{defrancesco06} \\
NGC4342 & S0 & 253 & $3.3 \times 10^8$ & $1.9 \times 10^8$ & $1.2 \times 10^{10}$ & \cite{haring04} \\
NGC4350 & S0 & 181 & $7.3 \times 10^8$ & $2.4 \times 10^8$ & $1.3 \times 10^{10}$ & \cite{pignatelli01} \\
NGC4459 & S0 & 178 & $0.7 \times 10^8$ & $1.3 \times 10^7$ & $7.9 \times 10^{10}$ & \cite{cappellari06} \\
NGC4552 & S0 & 252 & $4.8 \times 10^8$ & $0.8 \times 10^8$ & $1.9 \times 10^{11}$ & \cite{cappellari06} \\
NGC4564 & S0 & 157 & $0.6 \times 10^8$ & $0.3 \times 10^7$ & $4.4 \times 10^{10}$ & \cite{haring04} \\
NGC5128 & S0 & 120 & $0.5 \times 10^8$ & $1.8 \times 10^7$ & $3.6 \times 10^{10}$ & \cite{bekki03} \\
NGC5252 & S0 & 190 & $1.1 \times 10^9$ & $1.6 \times 10^9$ & $2.4 \times 10^{11}$ & \cite{marconi03} \\
NGC7332 & S0 & 135 & $1.3 \times 10^7$ & $0.6 \times 10^7$ & $1.5 \times 10^{10}$ & \cite{haring04} \\
NGC7457 & S0 & 69  & $3.5 \times 10^6$ & $1.4 \times 10^6$ & $7.0 \times 10^9$ & \cite{haring04} \\
NGC1023 & SB0& 204 & $4.4 \times 10^7$ & $0.5 \times 10^7$ & $6.9 \times 10^{10}$ & \cite{haring04} \\
NGC2778 & SB0& 162 & $1.4 \times 10^7$ & $0.9 \times 10^7$ & $1.1 \times 10^{10}$ & \cite{aller07} \\
NGC2787 & SB0& 210 & $4.1 \times 10^7$ & $0.5 \times 10^7$ & $2.9 \times 10^{10}$ & \cite{sarzi01} \\
NGC3384 & SB0& 148 & $1.6 \times 10^7$ & $0.2 \times 10^7$ & $2.0 \times 10^{10}$ & \cite{haring04} \\
NGC4596 & SB0& 149 & $7.9 \times 10^7$ & $3.8 \times 10^7$ & $2.6 \times 10^{10}$ & \cite{marconi03} \\
\noalign{\smallskip}\hline\noalign{\smallskip}
Circinus  & S  & 75  & $1.1 \times 10^6$ & $0.2 \times 10^6$ & $3.0 \times 10^9$ & \cite{hitschfeld08} \\
NGC224    & S  & 170 & $1.4 \times 10^8$ & $0.9 \times 10^8$ & $4.4 \times 10^{10}$ & \cite{riffeser08} \\
NGC1068   & S  & 151 & $8.4 \times 10^6$ & $0.3 \times 10^6$ & $1.5 \times 10^{10}$ & \cite{israel09} \\
NGC2748   & S  & 92  & $4.8 \times 10^7$ & $3.9 \times 10^7$ & $1.7 \times 10^{10}$ & \cite{atkinson05} \\
NGC3031   & S  & 162 & $7.6 \times 10^7$ & $2.2 \times 10^7$ & $1.0 \times 10^{10}$ & \cite{sofue98} \\
NGC4594   & S  & 240 & $1.0 \times 10^9$ & $1.0 \times 10^9$ & $2.7 \times 10^{11}$ & \cite{haring04} \\
Milky Way & SB & 100 & $3.7 \times 10^6$ & $0.2 \times 10^6$ & $1.1 \times 10^{10}$ & \cite{haring04} \\
NGC1300   & SB & 229 & $7.3 \times 10^7$ & $6.9 \times 10^7$ & $2.1 \times 10^{10}$ & \cite{atkinson05} \\
NGC3079   & SB & 146 & $2.4 \times 10^6$ & $2.4 \times 10^6$ & $1.7 \times 10^9$ & \cite{koda02} \\
NGC3227   & SB & 133 & $1.4 \times 10^7$ & $1.0 \times 10^7$ & $3.0 \times 10^9$ & \cite{wandel02} \\
NGC4151   & SB & 156 & $6.5 \times 10^7$ & $0.7 \times 10^7$ & $1.1 \times 10^{11}$ & \cite{wandel99} \\
NGC4258   & SB & 134 & $3.9 \times 10^7$ & $0.1 \times 10^7$ & $1.1 \times 10^{10}$ & \cite{marconi03} \\
NGC4945   & SB & 100 & $1.4 \times 10^6$ & $1.4 \times 10^6$ & $3.0 \times 10^9$ & \cite{hitschfeld08} \\
NGC7469   & SB & 153 & $1.2 \times 10^7$ & $1.4 \times 10^6$ & $4.5 \times 10^9$ & \cite{genzel95} \\
NGC7582   & SB & 156 & $5.5 \times 10^7$ & $2.6 \times 10^7$ & $1.3 \times 10^{11}$ & \cite{wold06} \\
\noalign{\smallskip}\hline
\end{tabular}
\end{table}
In Fig.~\ref{fig:1}(a)–-(c), we reported the $M_{\bullet}-\sigma$, $M_{\bullet}-M_{\mathrm{G}}$, and $M_{\bullet}-M_{\mathrm{G}}\sigma^2$ relations in log–-log plots (we associated a particular marker to each galaxy according to its morphological type). The best--fitting lines are also shown for each diagram. The comparison of the fits of the three relations reveals that the $\chi^2$, the intrinsic scatter $\varepsilon_{0}$, and the Pearson linear correlation coefficient $r$ of the $M_{\bullet}-M_{\mathrm{G}}\sigma^2$ relationship are better than the other ones (Table~\ref{tab:1}).
%

The lenticular galaxies NGC4342 and NGC4350 have been excluded from the fits because the SMBH mass to bulge mass ratio for these galaxies is very high~\cite{pignatelli01,bosch97}; their positions in Fig.~\ref{fig:1}(c) are enclosed in a dashed box. In Fig.~\ref{fig:1}(c) we also marked the position of the Milky Way and of the spiral NGC4594 (the Sombrero galaxy), which is surrounded by a halo of stars, dust, and gas that indicate it may actually be described as an elliptical galaxy that contains a more robust interior configuration~\cite{freeman07}. We refer the reader to~\cite{feoli09} for a comprehensive description of the diagrams reported in Fig.~\ref{fig:1}.

\subsection{G\"{u}ltekin et al. sample}
\label{sec:2.2}
In their catalog, G\"{u}ltekin et al. (2009) listed 55 galaxies, that had a direct measurement of the mass of their SMBHs, and gave the upper limits to SMBH masses of 18 additional galaxies; the latter are not considered in our analysis~\cite{gultekin09}. The values of the parameters of the 55 galaxies (27 ellipticals, 14 lenticulars, and 14 spirals) are listed in Table~\ref{tab:3}.
%
\begin{table}
\caption{The data are taken from Gultekin et al. (2009)~\cite{gultekin09}, except for the values of $M_{\mathrm{G}}$ whose references are given in the last column. The error for the bulge mass is 0.18 dex in the $\log_{10} {M_{\mathrm{G}}}$ for all galaxies.}
\label{tab:3}       
\begin{tabular}{lccccccc}
\hline\noalign{\smallskip}
Galaxy & Type & $\sigma_{\mathrm{e}} (km/s)$ & $M_{\mathrm{V}}$ & $M_{\bullet} (M_{\odot})$ & $\delta M_{\bullet} (M_{\odot})$ & $M_{\mathrm{G}} (M_{\odot})$ & References\\
\noalign{\smallskip}\hline\noalign{\smallskip}
A1836-BCG& E & $288 \pm 14$ & $-23.31 \pm 0.15$ & $3.9 \times 10^9$ & $0.6 \times 10^9$ & $7.6 \times 10^{11}$ & \cite{hu09} \\
A3565-BCG& E & $322 \pm 16$ & $-23.27 \pm 0.15$ & $5.2 \times 10^8$ & $0.8 \times 10^8$ & $6.1 \times 10^{11}$ & \cite{killen86} \\
CygnusA & E  & $270 \pm 14$ & $-21.27 \pm 0.10$ & $2.7 \times 10^9$ & $0.7 \times 10^9$ & $1.6 \times 10^{12}$ & \cite{marconi03} \\
IC1459  & E4 & $340 \pm 17$ & $-22.57 \pm 0.15$ & $2.8 \times 10^9$ & $1.2 \times 10^9$ & $6.6 \times 10^{11}$ & \cite{marconi03} \\
NGC221  & E2 & $ 75 \pm 3 $ & $-16.83 \pm 0.05$ & $3.1 \times 10^6$ & $0.6 \times 10^6$ & $8.0 \times 10^{8} $ & \cite{haring04} \\
NGC821  & E4 & $209 \pm 10$ & $-21.24 \pm 0.13$ & $4.2 \times 10^7$ & $3.5 \times 10^7$ & $1.3 \times 10^{11}$ & \cite{haring04} \\
NGC1399 & E1 & $337 \pm 16$ & $-22.13 \pm 0.10$ & $1.3 \times 10^9$ & $6.6 \times 10^8$ & $2.3 \times 10^{11}$ & \cite{houghton06} \\
NGC2778 & E2 & $175 \pm 8 $ & $-19.62 \pm 0.13$ & $1.6 \times 10^7$ & $1.0 \times 10^7$ & $1.1 \times 10^{10}$ & \cite{aller07} \\
NGC3377 & E6 & $145 \pm 7 $ & $-20.11 \pm 0.10$ & $1.1 \times 10^8$ & $1.1 \times 10^8$ & $3.1 \times 10^{10}$ & \cite{cappellari06} \\
NGC3379 & E0 & $206 \pm 10$ & $-21.10 \pm 0.03$ & $1.2 \times 10^8$ & $0.8 \times 10^8$ & $6.8 \times 10^{10}$ & \cite{haring04} \\
NGC3607 & E1 & $229 \pm 11$ & $-21.62 \pm 0.10$ & $1.2 \times 10^8$ & $0.4 \times 10^8$ & $1.6 \times 10^{11}$ & \cite{rickes09} \\
NGC3608 & E1 & $182 \pm 9 $ & $-21.05 \pm 0.10$ & $2.1 \times 10^8$ & $1.1 \times 10^8$ & $9.7 \times 10^{10}$ & \cite{haring04} \\
NGC4261 & E2 & $315 \pm 15$ & $-22.72 \pm 0.06$ & $5.5 \times 10^8$ & $1.2 \times 10^8$ & $3.6 \times 10^{11}$ & \cite{haring04} \\
NGC4291 & E2 & $242 \pm 12$ & $-20.67 \pm 0.13$ & $3.2 \times 10^8$ & $3.1 \times 10^8$ & $1.3 \times 10^{11}$ & \cite{haring04} \\
NGC4374 & E1 & $296 \pm 14$ & $-22.45 \pm 0.05$ & $1.5 \times 10^9$ & $1.1 \times 10^9$ & $3.6 \times 10^{11}$ & \cite{haring04} \\
NGC4459 & E2 & $167 \pm 8 $ & $-21.06 \pm 0.04$ & $7.4 \times 10^7$ & $1.4 \times 10^7$ & $7.9 \times 10^{10}$ & \cite{cappellari06} \\
NGC4473 & E4 & $190 \pm 9 $ & $-21.14 \pm 0.04$ & $1.3 \times 10^8$ & $0.9 \times 10^8$ & $9.2 \times 10^{10}$ & \cite{haring04} \\
NGC4486 & E1 & $375 \pm 18$ & $-22.92 \pm 0.04$ & $3.6 \times 10^9$ & $1.0 \times 10^9$ & $6.0 \times 10^{11}$ & \cite{haring04} \\
NGC4486A& E2 & $111 \pm 5 $ & $-18.70 \pm 0.05$ & $1.3 \times 10^7$ & $0.5 \times 10^7$ & $4.1 \times 10^{9} $ & \cite{nowak07} \\
NGC4649 & E2 & $385 \pm 19$ & $-22.65 \pm 0.05$ & $2.1 \times 10^9$ & $0.6 \times 10^9$ & $4.9 \times 10^{11}$ & \cite{haring04} \\
NGC4697 & E6 & $177 \pm 8 $ & $-21.29 \pm 0.11$ & $2.0 \times 10^8$ & $0.2 \times 10^8$ & $1.1 \times 10^{11}$ & \cite{haring04} \\
NGC4742 & E4 & $ 90 \pm 5 $ & $-19.91 \pm 0.10$ & $1.5 \times 10^7$ & $0.6 \times 10^7$ & $6.2 \times 10^{9} $ & \cite{haring04} \\
NGC5077 & E3 & $222 \pm 11$ & $-22.04 \pm 0.13$ & $8.0 \times 10^8$ & $5.0 \times 10^8$ & $2.1 \times 10^{11}$ & \cite{defrancesco08} \\
NGC5576 & E3 & $183 \pm 9 $ & $-21.26 \pm 0.13$ & $1.8 \times 10^8$ & $0.4 \times 10^8$ & $1.5 \times 10^{11}$ & \cite{gultekin09a} \\
NGC5845 & E3 & $234 \pm 11$ & $-19.77 \pm 0.13$ & $2.9 \times 10^8$ & $1.7 \times 10^8$ & $3.7 \times 10^{10}$ & \cite{haring04} \\
NGC6251 & E1 & $290 \pm 14$ & $-22.90 \pm 0.10$ & $6.0 \times 10^8$ & $2.0 \times 10^8$ & $5.6 \times 10^{11}$ & \cite{haring04} \\
NGC7052 & E3 & $266 \pm 13$ & $-22.35 \pm 0.10$ & $4.0 \times 10^8$ & $2.8 \times 10^8$ & $2.9 \times 10^{11}$ & \cite{haring04} \\
\noalign{\smallskip}\hline\noalign{\smallskip}
NGC3115 & S0 &  $230 \pm 11$ & $-21.18 \pm 0.05$ & $9.6 \times 10^8$ & $5.4 \times 10^8$ & $1.2 \times 10^{11}$ & \cite{haring04} \\
NGC3245 & S0 &  $205 \pm 10$ & $-20.96 \pm 0.10$ & $2.2 \times 10^8$ & $0.5 \times 10^8$ & $6.8 \times 10^{10}$ & \cite{haring04} \\
NGC3585 & S0 &  $213 \pm 10$ & $-21.80 \pm 0.20$ & $3.4 \times 10^8$ & $1.5 \times 10^8$ & $1.8 \times 10^{11}$ & \cite{gultekin09a} \\
NGC3998 & S0 &  $305 \pm 15$ & $-20.32 \pm 0.10$ & $2.4 \times 10^8$ & $2.1 \times 10^8$ & $5.5 \times 10^{10}$ & \cite{defrancesco06} \\
NGC4026 & S0 &  $180 \pm  9$ & $-19.83 \pm 0.20$ & $2.1 \times 10^8$ & $0.7 \times 10^8$ & $5.2 \times 10^{10}$ & \cite{haring04} \\
NGC4342 & S0 &  $225 \pm 11$ & $-18.84 \pm 0.10$ & $3.6 \times 10^8$ & $2.0 \times 10^8$ & $1.2 \times 10^{10}$ & \cite{haring04} \\
NGC4564 & S0 &  $162 \pm  8$ & $-19.60 \pm 0.32$ & $6.9 \times 10^7$ & $1.0 \times 10^7$ & $4.4 \times 10^{10}$ & \cite{haring04} \\
NGC5128 & S0/E& $150 \pm  7$ & $-21.82 \pm 0.08$ & $3.0 \times 10^8$ & $0.4 \times 10^8$ & $3.6 \times 10^{10}$ & \cite{bekki03} \\
NGC5252 & S0 &  $190 \pm 10$ & $\; \;\;\; \,...$ & $1.0 \times 10^9$ & $1.6 \times 10^9$ & $2.4 \times 10^{11}$ & \cite{marconi03} \\
NGC7457 & S0 &  $ 67 \pm  3$ & $-18.72 \pm 0.11$ & $4.1 \times 10^6$ & $1.7 \times 10^6$ & $7.0 \times 10^{9} $ & \cite{haring04} \\
NGC1023 & SB0 & $205 \pm 10$ & $-20.61 \pm 0.28$ & $4.6 \times 10^7$ & $0.5 \times 10^7$ & $6.9 \times 10^{10}$ & \cite{haring04} \\
NGC2787 & SB0 & $189 \pm  9$ & $-18.90 \pm 0.10$ & $4.3 \times 10^7$ & $0.5 \times 10^7$ & $2.9 \times 10^{10}$ & \cite{sarzi01} \\
NGC3384 & SB0 & $143 \pm  7$ & $-19.93 \pm 0.22$ & $1.8 \times 10^7$ & $0.3 \times 10^7$ & $2.0 \times 10^{10}$ & \cite{haring04} \\
NGC4596 & SB0 & $136 \pm  6$ & $-20.70 \pm 0.10$ & $8.4 \times 10^7$ & $4.4 \times 10^7$ & $2.6 \times 10^{10}$ & \cite{marconi03} \\
\noalign{\smallskip}\hline\noalign{\smallskip}
Circinus & S   & $158 \pm 18$ & $-17.36 \pm 0.10$ & $1.7 \times 10^6$ & $0.4 \times 10^6$ & $ 3.0 \times 10^{9} $ & \cite{hitschfeld08} \\
Milky Way& S   & $105 \pm 20$ & $\; \;\;\; \,...$ & $4.1 \times 10^6$ & $0.6 \times 10^6$ & $ 1.1 \times 10^{10}$ & \cite{haring04} \\
NGC224   & S   & $160 \pm  8$ & $-21.84 \pm 0.30$ & $1.5 \times 10^8$ & $0.9 \times 10^8$ & $ 4.4 \times 10^{10}$ & \cite{riffeser08} \\
NGC1068  & S   & $151 \pm  7$ & $-22.17 \pm 0.10$ & $8.6 \times 10^6$ & $0.3 \times 10^6$ & $ 1.5 \times 10^{10}$ & \cite{israel09} \\
NGC2748  & S   & $115 \pm  5$ & $-20.97 \pm 0.10$ & $4.7 \times 10^7$ & $3.8 \times 10^7$ & $ 1.7 \times 10^{10}$ & \cite{atkinson05} \\
NGC3031  & S   & $143 \pm  7$ & $-21.51 \pm 0.10$ & $8.0 \times 10^7$ & $6.9 \times 10^7$ & $ 1.0 \times 10^{10}$ & \cite{sofue98} \\
NGC4594  & S   & $240 \pm 12$ & $-22.44 \pm 0.15$ & $5.7 \times 10^8$ & $5.3 \times 10^8$ & $ 2.7 \times 10^{11}$ & \cite{haring04} \\
NGC4945  & S   & $134 \pm  7$ & $\; \;\;\; \,...$ & $1.4 \times 10^6$ & $0.7 \times 10^6$ & $ 3.0 \times 10^{9} $ & \cite{hitschfeld08} \\
NGC1300  & SB  & $218 \pm 10$ & $-21.34 \pm 0.10$ & $7.1 \times 10^7$ & $3.6 \times 10^7$ & $ 2.1 \times 10^{10}$ & \cite{atkinson05} \\
NGC3227  & SB  & $133 \pm 12$ & $-20.73 \pm 0.10$ & $1.5 \times 10^7$ & $0.8 \times 10^7$ & $ 3.0 \times 10^{9} $ & \cite{wandel02} \\
NGC4151  & SAB & $ 93 \pm  5$ & $-20.68 \pm 0.10$ & $4.5 \times 10^7$ & $0.5 \times 10^7$ & $ 1.1 \times 10^{11}$ & \cite{wandel99} \\
NGC4258  & SAB & $115 \pm 10$ & $-21.31 \pm 0.10$ & $3.8 \times 10^7$ & $0.1 \times 10^6$ & $ 1.1 \times 10^{10}$ & \cite{marconi03} \\
NGC4303  & SAB & $ 84 \pm  4$ & $-21.65 \pm 0.10$ & $4.5 \times 10^6$ & $9.5 \times 10^6$ & $ 1.6 \times 10^{9} $ & \cite{schinnerer02} \\
NGC7582  & SB  & $156 \pm 19$ & $-21.51 \pm 0.10$ & $5.5 \times 10^7$ & $1.6 \times 10^7$ & $ 1.3 \times 10^{11}$ & \cite{wold06} \\
\noalign{\smallskip}\hline
\end{tabular}
\end{table}
Again we consider that the error for the bulge mass is 0.18 dex in $\mathrm{log}_{10}{M_{\mathrm{G}}}$, whereas we use an absolute error of $\pm0.10$ mag for the visual magnitudes there where it is absent for some galaxies of the catalogue. We consider only 52 galaxies in the study of the $M_{\bullet}-L_{\mathrm{V}}$ relation, since the values of the magnitude of three of them have not been reported by G\"{u}ltekin et al. in their catalogue~\cite{gultekin09}. Moreover, these authors preferred to report the values of the effective dispersion velocity instead of the central one. However, as already noted by different authors~\cite{novak06,hu08}, the two ways of measuring the velocity dispersion do not generate profound differences. The V-band luminosities have been calculated from the extinction-corrected magnitudes, $M_{\mathrm{V,bulge}}$, using $\log_{10}\left({L_{\mathrm{V}}/L_{\odot,\mathrm{V}}}\right)=0.4\left(4.83-M_{\mathrm{V,bulge}}\right)$, in accordance with~\cite{gultekin09}.

\begin{figure}
\centering
\includegraphics[width=1.4\textwidth]{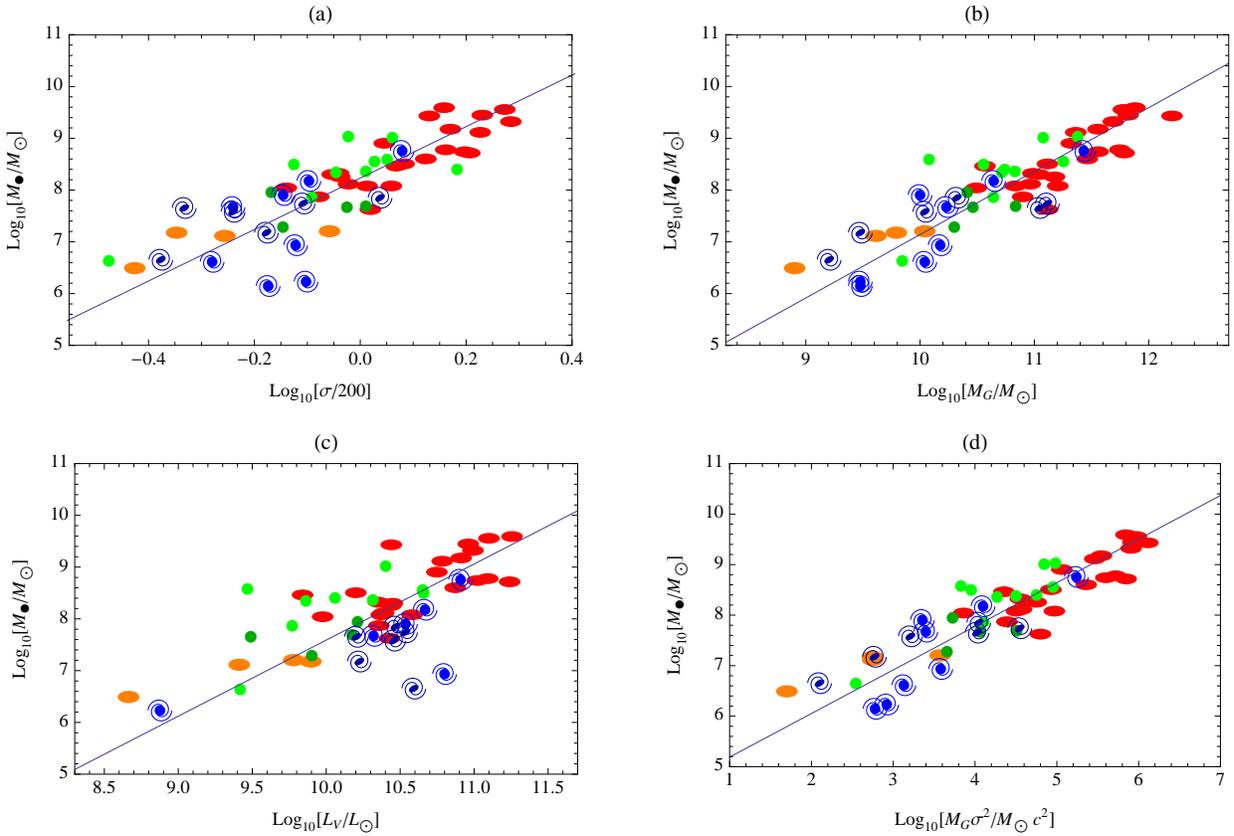}
\caption{Best-fitting (a) $M_{\bullet}-\sigma$, (b) $M_{\bullet}-M_{\mathrm{G}}$, (c) $M_{\bullet}-L_{\mathrm{V}}$, and (d) $M_{\bullet}-M_{\mathrm{G}}\sigma^2$ relations for the sample extracted from G\"{u}ltekin et al. (2009)~\cite{gultekin09}. The symbols are the same of Fig.~\ref{fig:1}.}
\label{fig:2}       
\end{figure}

The diagrams of the $M_{\bullet}-\sigma$, $M_{\bullet}-M_{\mathrm{G}}$, $M_{\bullet}-L_{\mathrm{V}}$, and $M_{\bullet}-M_{\mathrm{G}}\sigma^2$ relations are shown in log--log plots together with the best--fitting lines, see Fig.~\ref{fig:2}(a)–-(d) (the symbols are the same of Fig.~\ref{fig:1}). From inspection of Table~\ref{tab:1}, it is clear that the $M_{\bullet}-M_{\mathrm{G}}\sigma^2$ relation works better than the other relations when applied on the same sample of galaxies.

%
%

\subsection{Hu sample}
\label{sec:2.3}
Hu (2009) compiled a catalogue of 58 galaxies (28 ellipticals, 19 lenticulars, and 11 spirals)~\cite{hu09}. The main parameters are summarized in Table~\ref{tab:4}. He estimated the dynamical mass of bulges by $M_{\mathrm{G,dyn}}=k R_{\mathrm{e}}\sigma_{\mathrm{c}}^2/G$, where $R_{\mathrm{e}}$ is the bulge effective radius, $G$ is the gravitational constant, and $k$ is a function of the S\'{e}rsic index which is determined numerically.

Of course, there are also other methods to estimate the mass of the galaxies, and in particular we prefer those which use the Jeans equation or three--dimensional models, see appendix B of~\cite{feoli09}.

The $1\sigma$ errors of $L_{\mathrm{K}}$ and $\sigma_{\mathrm{c}}$ are adopted as $10\%$ and $5\%$ respectively, whereas that of $\log{M_{\mathrm{G}}}$ is adopted as 0.15 dex $(40\%)$~\cite{hu09}.

\begin{figure}
\centering
\includegraphics[width=1.4\textwidth]{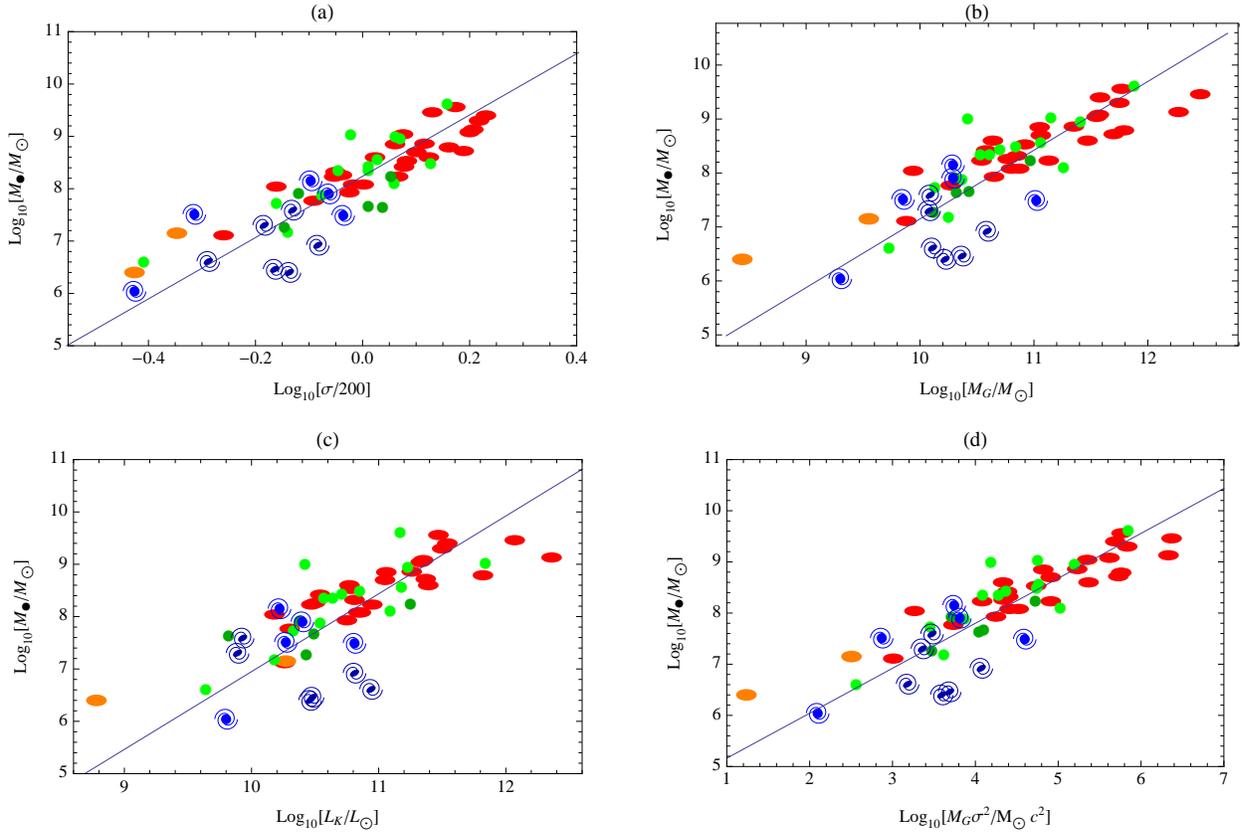}
\caption{Best-fitting (a) $M_{\bullet}-\sigma$, (b) $M_{\bullet}-M_{\mathrm{G}}$, (c) $M_{\bullet}-L_{\mathrm{K}}$, and (d) $M_{\bullet}-M_{\mathrm{G}}\sigma^2$ relations for the sample extracted from Hu (2009)~\cite{hu09}. The symbols are the same of Fig.~\ref{fig:1}.}
\label{fig:3}       
\end{figure}

\begin{table}
\caption{The data are all taken from Hu (2009)~\cite{hu09}. The $1\sigma$ errors of $L_{\mathrm{K}}$ and $\sigma_{\mathrm{c}}$ are $10\%$ and $5\%$ respectively, whereas that of $\log{M_{\mathrm{G}}}$ are 0.15 dex.}
\label{tab:4}       
\begin{tabular}{lcccccc}
\hline\noalign{\smallskip}
Galaxy & Type & $\sigma_{\mathrm{e}} (km/s)$ & $\mathrm{Log}_{10}\left[L_{\mathrm{K}}/L_{\odot,\mathrm{K}}\right]$ & $M_{\bullet} (M_{\odot})$ & $\delta M_{\bullet} (M_{\odot})$ & $M_{\mathrm{G}} (M_{\odot})$ \\
\noalign{\smallskip}\hline\noalign{\smallskip}
NGC821   & E4 & 189 & 10.75 & $8.5 \times 10^7$ & $5.9 \times 10^7$ & $4.5 \times 10^{10}$ \\
NGC1399  & E1 & 317 & 11.35 & $1.2 \times 10^9$ & $1.2 \times 10^9$ & $3.7 \times 10^{11}$ \\
NGC2974  & E4 & 233 & 10.95 & $1.7 \times 10^8$ & $2.1 \times 10^7$ & $1.3 \times 10^{11}$ \\
NGC3377  & E5 & 138 & 10.18 & $1.1 \times 10^8$ & $1.1 \times 10^8$ & $8.7 \times 10^{9}$ \\
NGC3379  & E1 & 201 & 10.86 & $1.2 \times 10^8$ & $1.2 \times 10^8$ & $7.4 \times 10^{10}$ \\
NGC3608  & E2 & 178 & 10.81 & $2.1 \times 10^8$ & $1.1 \times 10^8$ & $7.1 \times 10^{10}$ \\
NGC4473  & E5 & 192 & 10.84 & $1.2 \times 10^8$ & $3.2 \times 10^8$ & $6.3 \times 10^{10}$ \\
NGC4486  & E0 & 298 & 11.47 & $3.6 \times 10^9$ & $1.4 \times 10^9$ & $5.9 \times 10^{11}$ \\
NGC4486A & E2 & 110 & 10.26 & $1.3 \times 10^7$ & $1.5 \times 10^7$ & $7.6 \times 10^{9}$ \\
NGC4552  & E  & 252 & 11.05 & $5.0 \times 10^8$ & $6.1 \times 10^7$ & $1.1 \times 10^{11}$ \\
NGC4564  & E6 & 162 & 10.30 & $5.9 \times 10^7$ & $8.7 \times 10^6$ & $1.8 \times 10^{10}$ \\
NGC4261  & E2 & 309 & 11.37 & $5.2 \times 10^8$ & $1.4 \times 10^8$ & $5.0 \times 10^{11}$ \\
NGC4291  & E2 & 242 & 10.78 & $3.4 \times 10^8$ & $9.8 \times 10^8$ & $8.3 \times 10^{10}$ \\
NGC4621  & E5 & 211 & 10.77 & $4.0 \times 10^8$ & $4.9 \times 10^7$ & $4.4 \times 10^{10}$ \\
NGC4649  & E1 & 330 & 11.50 & $2.0 \times 10^9$ & $8.2 \times 10^8$ & $5.6 \times 10^{11}$ \\
NGC4697  & E4 & 177 & 10.47 & $1.7 \times 10^8$ & $3.4 \times 10^7$ & $3.5 \times 10^{10}$ \\
NGC5077  & E3 & 261 & 11.26 & $7.2 \times 10^8$ & $5.1 \times 10^8$ & $2.2 \times 10^{11}$ \\
NGC5576  & E3 & 183 & 10.50 & $1.8 \times 10^8$ & $5.2 \times 10^7$ & $5.9 \times 10^{10}$ \\
NGC5813  & E1 & 230 & 11.06 & $7.1 \times 10^8$ & $8.6 \times 10^7$ & $1.1 \times 10^{11}$ \\
NGC5845  & E3 & 239 & 10.54 & $2.6 \times 10^8$ & $3.7 \times 10^8$ & $3.8 \times 10^{10}$ \\
NGC5846  & E0 & 238 & 11.33 & $1.1 \times 10^9$ & $1.1 \times 10^8$ & $3.5 \times 10^{11}$ \\
NGC6251  & E2 & 290 & 11.82 & $6.2 \times 10^8$ & $3.2 \times 10^8$ & $6.2 \times 10^{11}$ \\
NGC7052  & E4 & 266 & 11.39 & $4.0 \times 10^8$ & $2.8 \times 10^8$ & $3.0 \times 10^{11}$ \\
IC4296   & E  & 322 & 12.36 & $1.3 \times 10^9$ & $2.4 \times 10^8$ & $1.9 \times 10^{12}$ \\
CygnusA  & E  & 270 & 12.07 & $2.9 \times 10^9$ & $9.2 \times 10^8$ & $2.9 \times 10^{12}$ \\
IC1459   & E3 & 340 & 11.54 & $2.5 \times 10^9$ & $5.1 \times 10^8$ & $3.8 \times 10^{11}$ \\
NGC221   & E2 &  75 & 8.780 & $2.5 \times 10^6$ & $6.5 \times 10^6$ & $2.8 \times 10^{8}$ \\
NGC4742  & E4 &  90 & 10.27 & $1.4 \times 10^7$ & $7.8 \times 10^6$ & $3.5 \times 10^{9}$ \\
\noalign{\smallskip}\hline\noalign{\smallskip}
NGC524   & S0 & 235 & 11.23 & $8.3 \times 10^8$ & $5.9 \times 10^7$ & $2.6 \times 10^{11}$ \\
NGC2549  & S0 & 145 & 10.18 & $1.4 \times 10^7$ & $5.2 \times 10^7$ & $1.8 \times 10^{10}$ \\
NGC3115  & S0 & 230 & 10.42 & $9.3 \times 10^8$ & $5.5 \times 10^8$ & $2.6 \times 10^{10}$ \\
NGC3245  & S0 & 205 & 10.64 & $2.1 \times 10^8$ & $6.6 \times 10^7$ & $4.1 \times 10^{10}$ \\
NGC3414  & S0 & 205 & 10.71 & $2.5 \times 10^8$ & $3.7 \times 10^7$ & $5.0 \times 10^{10}$ \\
NGC3585  & S0 & 213 & 11.18 & $3.4 \times 10^8$ & $1.5 \times 10^8$ & $1.1 \times 10^{11}$ \\
NGC3607  & S0 & 229 & 11.09 & $1.2 \times 10^8$ & $6.2 \times 10^7$ & $1.8 \times 10^{11}$ \\
NGC3998  & S0 & 268 & 10.85 & $2.9 \times 10^8$ & $5.8 \times 10^7$ & $6.9 \times 10^{10}$ \\
NGC4026  & S0 & 180 & 10.57 & $2.1 \times 10^8$ & $6.6 \times 10^7$ & $3.4 \times 10^{10}$ \\
NGC4459  & S0 & 168 & 10.54 & $7.1 \times 10^7$ & $1.6 \times 10^7$ & $2.3 \times 10^{10}$ \\
NGC5128  & S0 & 138 & 10.33 & $5.0 \times 10^7$ & $6.1 \times 10^6$ & $1.3 \times 10^{10}$ \\
NGC5252  & S0 & 190 & 11.84 & $1.0 \times 10^9$ & $1.5 \times 10^9$ & $1.4 \times 10^{11}$ \\
NGC7457  & S0 &  78 &  9.64 & $3.8 \times 10^6$ & $1.7 \times 10^6$ & $5.4 \times 10^{9}$ \\
P49940   & S0 & 288 & 11.17 & $3.9 \times 10^9$ & $5.7 \times 10^8$ & $7.6 \times 10^{11}$ \\
NGC1023  & SB0& 205 & 10.49 & $4.4 \times 10^7$ & $5.3 \times 10^6$ & $2.7 \times 10^{10}$ \\
NGC1316  & SB0& 226 & 11.25 & $1.6 \times 10^8$ & $3.3 \times 10^7$ & $9.3 \times 10^{10}$ \\
NGC4596  & SB0& 152 & 10.38 & $7.8 \times 10^7$ & $5.7 \times 10^7$ & $2.0 \times 10^{10}$ \\
NGC2787  & SB0& 218 &  9.82 & $4.1 \times 10^7$ & $6.0 \times 10^6$ & $2.1 \times 10^{10}$ \\
NGC3384  & SB0& 143 & 10.43 & $1.7 \times 10^7$ & $2.6 \times 10^6$ & $1.3 \times 10^{10}$ \\
\noalign{\smallskip}\hline\noalign{\smallskip}
Circinus & S  &  75 &  9.80 & $1.1 \times 10^6$ & $2.5 \times 10^6$ & $2.0 \times 10^{9}$ \\
NGC3393  & S  & 184 & 10.81 & $3.1 \times 10^7$ & $2.2 \times 10^6$ & $1.0 \times 10^{11}$ \\
NGC224   & S  & 160 & 10.22 & $1.4 \times 10^8$ & $9.3 \times 10^7$ & $1.9 \times 10^{10}$ \\
NGC3031  & S  & 173 & 10.40 & $7.9 \times 10^7$ & $2.1 \times 10^7$ & $1.9 \times 10^{10}$ \\
NGC4151  & S  &  97 & 10.27 & $3.2 \times 10^7$ & $8.2 \times 10^7$ & $7.1 \times 10^{9}$ \\
IC2560   & SB & 137 & 10.48 & $2.9 \times 10^6$ & $7.5 \times 10^6$ & $2.3 \times 10^{10}$ \\
Milky Way& SB & 103 & 10.25 & $4.1 \times 10^6$ & $0.6 \times 10^6$ & $1.3 \times 10^{10}$ \\
NGC1068  & SB & 165 & 10.81 & $8.3 \times 10^6$ & $3.9 \times 10^6$ & $3.9 \times 10^{10}$ \\
NGC3079  & SB & 146 & 10.46 & $2.5 \times 10^6$ & $2.5 \times 10^6$ & $1.7 \times 10^{10}$ \\
NGC3227  & SB & 131 &  9.89 & $2.0 \times 10^7$ & $3.8 \times 10^7$ & $1.2 \times 10^{10}$ \\
NGC4258  & SB & 148 &  9.93 & $3.9 \times 10^7$ & $9.1 \times 10^6$ & $1.2 \times 10^{10}$ \\
\noalign{\smallskip}\hline
\end{tabular}
\end{table}
In Fig.~\ref{fig:3}(a)–-(d), we show the $M_{\bullet}-\sigma$, $M_{\bullet}-M_{\mathrm{G}}$, $M_{\bullet}-L_{\mathrm{K}}$, and $M_{\bullet}-M_{\mathrm{G}}\sigma^2$ relations in log–-log plots (the symbols are the same as those used in Fig.~\ref{fig:1}), together with the best--fitting lines.
%
We recovered the same result of~\cite{hu09}, that is the intrinsic scatter of the $M_{\bullet}-L_{\mathrm{K}}$ relation is still larger than that of the $M_{\bullet}-\sigma$ relation, while the $M_{\bullet}-M_{\mathrm{G}}$ relation is as tight as the $M_{\bullet}-\sigma$ relation. However, the $M_{\bullet}-M_{\mathrm{G}}\sigma^2$ relation undoubtedly gives the best fit of the experimental data, see Table~\ref{tab:1}. This means that $M_{\mathrm{G}}\sigma^2$ is a better tracer of $M_{\bullet}$ than $L_{\mathrm{K}}$, $M_{\mathrm{G}}$, and $\sigma$.

\section{Age--temperature diagram for galaxies}
\label{sec:3}
The main finding that emerges from the analysis of the data is that the values of the intrinsic scatter and of the $\chi^2$ of the relation between the mass of SMBH located in the center of the host galaxies and the kinetic energy of random motions of the corresponding bulges are better than those of the other relations. This statement is true for each of the three samples considered in this work, which have been extracted from independent catalogues. Besides having a support from numerical models~\cite{hopkins07b,marulli08}, the $M_{\bullet}-M_{\mathrm{G}}\sigma^2$ law is also the only one with a quite clear physical explanation. In fact, the mass of the central BH, just like the entropy, can only increase with time or at most remain the same but never decrease. So, $M_{\bullet}$ is connected with the age of the galaxy. On the other side, the kinetic energy of the stellar bulges directly determines the temperature of the galactic system~\cite{feoli09}. Hence, it is possible to reinterpret our relation as an age--temperature diagram for galaxies.

There is also an evident indication that the galaxies occupy quite well--defined regions of the $M_{\bullet}-(M_{\mathrm{G}}\sigma^2)/c^2$ plane, accordingly to their morphological type. In fact, they are distributed along the best--fitting line, in such a way that the ellipticals are in the upper part of the diagram, the lenticulars in the middle, and the spirals in the lower part, see Fig.~\ref{fig:1}(c), Fig.~\ref{fig:2}(d), and Fig.~\ref{fig:3}(d). The distribution of the galaxy types changes a little bit moving from one sample to another depending on the differences of the morphological classification of some galaxies given by the various authors of the catalogues. This is essentially caused by the fact that whenever the lenticulars are inclined face--on it is often difficult to distinguish between them and ellipticals. On the other hand, many of the flattest ellipticals may actually be misclassified lenticulars~\cite{vandenbergh09a}. Actually, most of the lenticulars are spread and mixed with the low/intermediate-mass ellipticals, supporting the emerging idea that their nature is different from the standard one that sees them just as a transition type between ellipticals and spirals~\cite{vandenbergh09b}. The barred lenticulars are generally located in a lower zone than that occupied by lenticulars. The few galaxies of types E4 and E6 appear only in the middle region, this because the mass of their SMBH is roughly an order of magnitude lower than than those in host galaxies of types E0 and E1. Finally, spirals and barred spirals are located in the lower zone of the FM diagram, even if their distribution above and under the best--fitting line is not clear since it changes dramatically depending on the sample considered.


Together with the spirals, we find some dwarf elliptical galaxies (dEs) in the lower part of the diagram, that supports the hypothesis that the dEs could have suffered a morphology changing due to a process called ``galaxy harassment''~\cite{moore96}. To be more precise, the dEs could be the remnants of low-mass spiral galaxies that obtained a rounder shape through the action of repeated gravitational interactions with giant galaxies within a cluster. If this theory is exact, than it is possible to explain the presence of the dEs in the zone of the FM diagram where most of the late type galaxies are confined. On the other hand, the idea that the dEs may be primordial objects, rather than young galaxies, is in part supported by numerical simulations, which predict a very low SMBH mass limit for the ellipticals, smaller with respect to observations, as is discussed in the next section. This hypothesis could be correct if new ellipticals will be found in the middle region of the diagram.


For early type (E + S0) galaxies, the ratio of the galaxy mass to SMBH mass is essentially independent of galaxy flattening, i.e. the fraction of the total mass of early-type galaxies that is in the form of a central BH is independent of  the angular momentum of the host galaxy. In fact, using the data of the galaxy sample of Curir et al.~\cite{curir93}, the correlation coefficient of the relation between the ratio $(M_{\bullet}/M_{\mathrm{G}})$ and the specific angular momentum is very low, $r=0.54$. This agrees with the fact that the kinetic energy of stellar random motions of the early type galaxies is higher than that of the late type galaxies (see the differences between the values of $\sigma$ in Tables~\ref{tab:2},~\ref{tab:3}, and~\ref{tab:4}). However, it is puzzling that the lenticular galaxies NGC4342 and NGC4350 present a high values of the SMBH mass to bulge mass ratio, see Fig.~\ref{fig:1}(c). Their distance from the best--fitting line is quite suspicious and maybe is an indication of a peculiar evolutionary story.

\section{Numerical models}
\label{sec:4}
We compare the observed age-temperature relation described in the previous sections with the predictions of two hierarchical models of structure formation, one by De Lucia \& Blaizot 2007~\cite{delucia07} (the `Max Planck Institut f\"{u}r Astrophysik (MPA) model') and the other by Bower et
al. 2006~\cite{bower06} (the `Durham model'), that simulate the cosmological co-evolution of dark matter haloes, subhaloes, galaxies and SMBH in the $\Lambda$CDM--cosmology framework. Both models have been implemented on top of a high-resolution cosmological N-body simulation, the Millennium Run~\cite{springel05}, that follows the evolution of $N= 2160^3$ dark matter particles of mass $8.6\times10^{8}\,h^{-1}{\rm M}_{\odot}$, within a co-moving box of size $500\, h^{-1}$Mpc on a side and scale resolution of 5 $h^{-1}\,\mbox{kpc}\,$, from $z=127$ to the present. The cosmological parameters adopted are $\Omega_{\rm  m}=0.25$, $\Omega_{\rm b}=0.045$, $h=0.73$, $\Omega_\Lambda=0.75$, $n=1$, and $\sigma_8=0.9$~\cite{spergel03}. The output of both models, dubbed {\em DeLucia2006a} and {\em Bower2006a}, respectively, are publicly available at http://www.mpa-garching.mpg.de/millennium~\cite{lemson06}.

\begin{figure}
\includegraphics[width=1.4\textwidth]{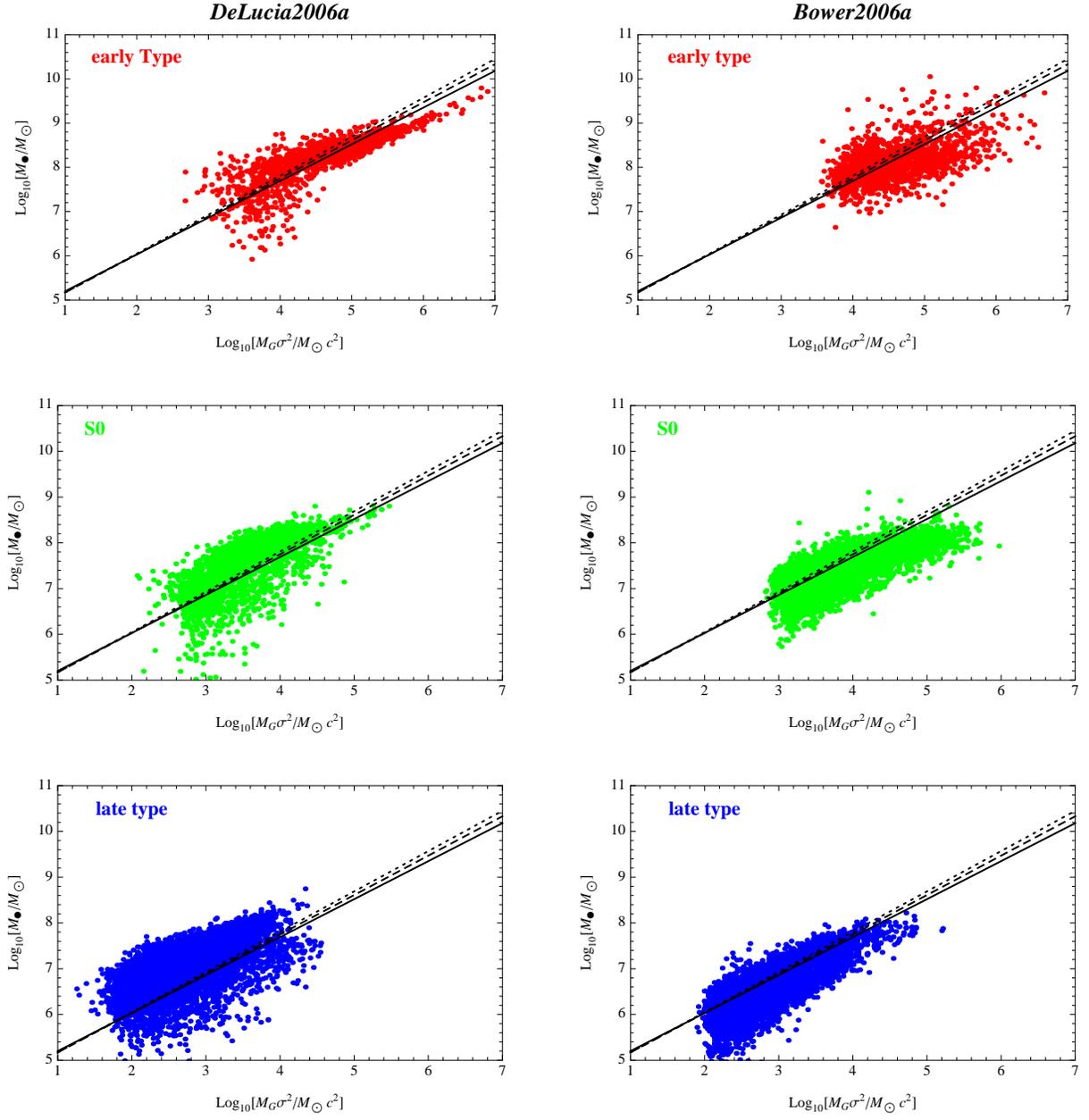}
\caption{The local $M_{\bullet}-M_{\mathrm{G}} \sigma^2$
  relation predicted by the two cosmological models considered. Red,
  green and blue dots refer to early type, S0 and late type galaxies,
  respectively. The black lines show the linear best fit relations to
  the observational dataset, reported in Table~\ref{tab:1}: Graham
  2008 sample~\cite{graham08} - solid line, G\"{u}ltekin et al. 2009
  sample~\cite{gultekin09} - dashed line and Hu 2009~\cite{hu09}
  sample - dotted lines. Only $1/100$ of the total galaxy sample is
  shown in the figure.}
\label{fig:4}
\end{figure}

\begin{figure}
\includegraphics[width=0.8\textwidth]{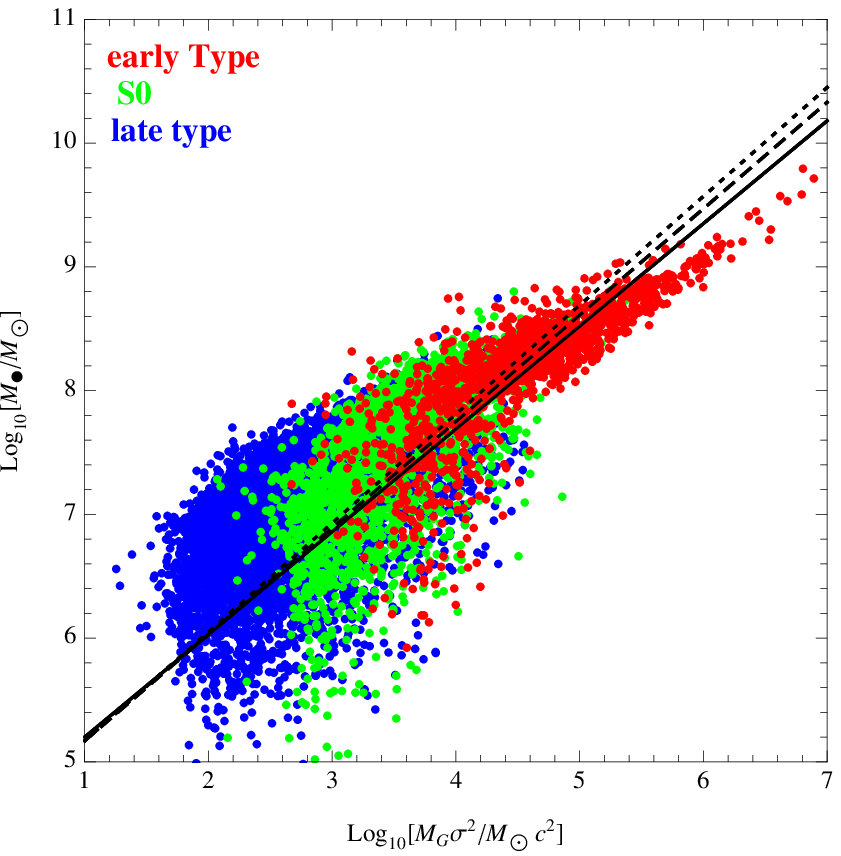}
\caption{The local $M_{\bullet}-M_{\mathrm{G}} \sigma^2$ relation
  predicted by the MPA model in a compact form, to directly compare
  with Fig.~\ref{fig:1}(c), ~\ref{fig:2}(c) and ~\ref{fig:3}(c). The
  symbols are as is Fig.~\ref{fig:4}.}
\label{fig:5}
\end{figure}

The high mass and spatial resolution of the Millennium Simulation allows one to track the motion of dark matter substructures inside massive haloes, making it possible to construct merging history trees of all the dark matter haloes and subhaloes inside the simulation box~\cite{springel05}. The dynamical evolution of all satellite galaxies is followed until tidal truncation and stripping disrupt their host dark matter subhaloes.  Then a residual survival time is estimated by computing the dynamical friction formula.  To populate the dark matter subhaloes with galaxies and black holes, both models adopt a set of equations to describe the radiative cooling of gas, the star formation, the metal enrichment and supernovae feedback, the growth and feedback of SMBH, the UV background reionization, and the effects of galaxy mergers.

In the MPA model, the BH mass accretion is triggered by two different phenomena: i) the merger between gas-rich galaxies ({\em quasar mode}) and ii) the cooling flow at the centers of X-ray emitting atmospheres in galaxy groups and clusters ({\em radio mode}). In the Durham model, in addition to these two mechanisms, the BH mass accretion is also triggered by the disk instability. The most important consequence of the BH growth in the evolution of galaxies is that the energy feedback associated with the {\em radio mode} accretion is assumed to be able to reduce or even stop the cooling flow in the halo centers.

A full description of the MPA and Durham models and a comparison between their main predictions with observations can be found in~\cite{croton06,delucia07,marulli08,bonoli09,marulli09} and in~\cite{bower06,malbon07}, respectively.

In both models the morphology of a galaxy is determined by the bulge-to-disc ratio of its absolute, rest frame B-band luminosity. Specifically, a galaxy is classified as early type if $\Delta M<0.4$, as lenticular (S0) if $0.4<\Delta M<1.56$ and as late type if $\Delta M>1.56$, where $\Delta M=M_{\rm bulge}-M_{\rm total}$ and $M_{\rm bulge}$ and $M_{\rm total}$ are the B-band magnitude of the bulge and of the whole galaxy, respectively~\cite{delucia07}.

In Fig.~\ref{fig:4} we show the local ($z=0$) $M_{\bullet}-M_{\mathrm{G}} \sigma^2$ relation predicted by the two models described above. As explicitly indicated by the labels, red, green and blue dots refer to early type, S0 and late type galaxies, respectively. The bulge velocity dispersion, $\sigma$, is derived from the ${\rm V}_{\rm c}-\sigma$ relation of Baes et al. 2003~\cite{baes03}, as in~\cite{marulli08}.  To closely mimic the observed catalogues, we have selected the model galaxies in these three observed mass ranges: $3.7\times10^{10}<M_{\mathrm{G}}/M_\odot<1.6\times10^{12}$ for early type, $10^{10}<M_{\mathrm{G}}/M_\odot<2.4\times10^{11}$ for S0, and $1.7\times10^9<M_{\mathrm{G}}/M_\odot<1.3\times10^{11}$ for late type galaxies. Besides, for clarity reasons, we have shown only $1/100$ of the total galaxy sample. The black lines show the linear best fit relations to the observational dataset, reported in Table~\ref{tab:1}: Graham 2008 sample~\cite{graham08} - solid line, G\"{u}ltekin et al. 2009 sample~\cite{gultekin09} - dashed line and Hu 2009~\cite{hu09} sample - dotted lines. As can be seen in Fig.~\ref{fig:4}, the predictions of both the MPA and Durham model are in quite good agreement with the observed scaling relations.
Also, the upper BH mass limits of the three groups considered (early type, S0 and late type) match the observed ones, while the lower BH mass limits predicted are smaller with respect to observations.  More accurate observational data are necessary to understand if this is a problem of the models themselves or simply depends on observational selection effects due to the fact that the number of galaxies with a realiable measured SMBH mass is still low.

Finally, to directly compare the MPA model predictions with the observational data reported in Fig.~\ref{fig:1}(c), ~\ref{fig:2}(d) and~\ref{fig:3}(d), we have shown in Fig.~\ref{fig:5} the $M_{\bullet}-M_{\mathrm{G}} \sigma^2$ relation in a more compact form.

\section{Summary}
\label{sec:5}
We have tested the goodness of the $M_{\bullet}-M_{\mathrm{G}} \sigma^2$ relation as a predictor of the SMBH mass in the center of galaxies on three different samples based on the galaxy catalogues of Graham~\cite{graham08}, G\"{u}ltekin et al.~\cite{gultekin09}, and Hu~\cite{hu09}. Our analysis shows that this relation is the one with the lowest scatter when compared with other relations like $M_{\bullet}-\sigma$, $M_{\bullet}-M_{\mathrm{G}}$, and $M_{\bullet}-L_{\mathrm{G}}$. This is evident if we look into the figures~\ref{fig:1},~\ref{fig:2}, and~\ref{fig:3}, and the Table~\ref{tab:1}, where the values of the intrinsic scatter and of the $\chi^2$ are reported, providing a quick comparison among the various relations. The galaxies arrange themselves on the $M_{\bullet}-M_{\mathrm{G}} \sigma^2$ plane following the best--fitting line and in accordance with their morphological type. In particular, the late types are clearly separated from the early types; the lenticulars are more mixed with the ellipticals; the barred lenticulars are more mixed with the spirals; the dwarf ellipticals occupy the same region of the spirals possibly revealing themselves as remnants of low--mass spirals, see Fig.~\ref{fig:1}(c), ~\ref{fig:2}(d) and~\ref{fig:3}(d). From this point of view, the $M_{\bullet}-M_{\mathrm{G}} \sigma^2$ relation can be revised as an age-temperature diagram for the galaxies, where the age is deducted from the mass of the central SMBHs, while the temperature from the kinetic energy of the random motions of the stars belonging to the spheroidal components~\cite{feoli09}.

The soundness of the FM diagram has also been investigated in the $\Lambda$CDM cosmology using two galaxy formation models based on the Millennium Simulation, one by Bower et al. (the Durham model)~\cite{bower06} and the other by De Lucia \& Blaizot (the MPA model)~\cite{delucia07}. Both the MPA and the Durham model reproduce quite well the real data, taking into account the uncertainties of the present–day observational measurements.


\begin{acknowledgements}
L.M. acknowledges support for this work by research funds of the University of Salerno and the Italian Space Agency (ASI), and A.F. by research funds of the University of Sannio.
\end{acknowledgements}


\begin{thebibliography}{}
%
\bibitem{kormendy95}
Kormendy, J., Richstone, D.: ARA\&A \textbf{33}, 581 (1995)
%
\bibitem{richstone98}
Richstone, D., Ajhar, E.~A., Bender, R., et al.: Nature \textbf{395}, A14 (1998)
%
\bibitem{magorrian98}
Magorrian, J., Tremaine, S., Richstone, D., et al.: AJ \textbf{115}, 2285 (1998)
%
\bibitem{laor01}
Laor, A.: ApJ \textbf{553}, 677 (2001)
%
\bibitem{wandel02}
Wandel, A.: ApJ \textbf{565}, 762 (2002)
%
\bibitem{marconi03}
Marconi, A., Hunt, L.~K.: ApJ \textbf{589}, L21 (2003)
%
\bibitem{haring04}
H\"{a}ring, N., Rix, H.: ApJL \textbf{604}, L89 (2004)
%
\bibitem{ferrarese00}
Ferrarese, L., Merritt, D.: ApJ \textbf{539}, L9 (2000)
%
\bibitem{gebhardt00}
Gebhardt, K., Bender, R., Bower, G., et al.: ApJL \textbf{539}, 13 (2000)
%
\bibitem{tremaine02}
Tremaine, S., Gebhardt, K., Bender, R., et al.: ApJ \textbf{574}, 740 (2002)
%
\bibitem{feoli05}
Feoli, A., Mele, D.: Int. Jour. Mod. Phys. D \textbf{14}, 1861 (2005)
%
\bibitem{feoli07}
Feoli, A., Mele, D.: Int. Jour. Mod. Phys. D \textbf{16}, 1261 (2007)
%
\bibitem{graham05}
Graham, A.~W., Driver, S.~P.: PASA \textbf{22}, 118 (2005)
%
\bibitem{graham07}
Graham, A.~W., Driver, S.~P.: ApJ \textbf{655}, 77 (2007)
%
\bibitem{novak06}
Novak, G.~S., Faber, S.~M., Dekel, A.: ApJ \textbf{637}, 96 (2006)
%
\bibitem{hopkins07a}
Hopkins, P.~F., Hernquist, L., Cox, T. J., et al.: ApJ \textbf{669}, 67 (2007)
%
\bibitem{hopkins07b}
Hopkins, P.~F., Hernquist, L., Cox, T. J., et al.: ApJ \textbf{669}, 45 (2007)
%
\bibitem{marulli08}
Marulli, F., Bonoli, S., Branchini, E., et al.: MNRAS \textbf{385}, 1846 (2008)
%
\bibitem{silk98}
Silk, J., Rees, M.~J.: A\&A \textbf{331}, L1 (1998)
%
\bibitem{ferrarese02}
Ferrarese, L.: ApJ \textbf{578}, 90 (2002)
%
\bibitem{booth09}
Booth, C.~M., Schaye, J.: MNRAS submitted (2009)
%
\bibitem{kollmeier06}
Kollmeier, J.~A., Onken, C.~A., Kochanek, C.~S.: ApJ \textbf{648}, 128 (2006)
%
\bibitem{merloni03}
Merloni, A., Heinz, S., Di Matteo, T.: MNRAS, \textbf{345}, 1057 (2003)
%
\bibitem{falcke04}
Falcke, H., K\"{o}rding, E., Markoff, S.: A\&A \textbf{414}, 895 (2004)
%
\bibitem{gultekin09b}
G\"{u}ltekin, K., Cackett, E. M., Miller, J.~M., et al.: ApJ \textbf{706}, 404 (2009)
%
\bibitem{feoli09}
Feoli, A., Mancini, L.: ApJ \textbf{703}, 1502 (2009)
%
\bibitem{graham08}
Graham, A.~W.: PASA \textbf{25}, 167 (2008)
%
\bibitem{gultekin09}
G\"{u}ltekin, K., Richstone, D. O., Gebhardt, K., et al.: ApJ \textbf{698}, 198 (2009)
%
\bibitem{hu09}
Hu, J.: MNRAS submitted (2009)
%
\bibitem{press92}
Press, W.~H., Teukolsky, S.~A., Vetterling, W.~T., Flannery, B.~P.: Numerical Recipes (2nd ed.; Cambridge: Cambridge Univ. Press) (1992)
%
\bibitem{dallabonta07}
Dalla Bont\`{a}, E., Ferrarese, L., Corsini, E.~M., et al.: Mem. S.A.It. \textbf{78}, 745 (2007)
%
\bibitem{houghton06}
Houghton, R.~C.~W., Magorrian, J., Sarzi, M., et al.: MNRAS \textbf{367}, 2 (2006)
%
\bibitem{cappellari06}
Cappellari, M., Bacon, R., Bureau, M., et al.: MNRAS \textbf{366}, 1126 (2006)
%
\bibitem{nowak07}
Nowak, N., Saglia, R.~P., Thomas, J., et al.: MNRAS \textbf{379}, 909 (2007)
%
\bibitem{defrancesco08}
De Francesco, G., Capetti, A., Marconi, A.: A\&A \textbf{479}, 355 (2008)
%
\bibitem{defrancesco06}
De Francesco, G., Capetti, A., Marconi, A.: A\&A \textbf{460}, 439 (2006)
%
\bibitem{pignatelli01}
Pignatelli, E., Salucci, P., Danese, L.: MNRAS \textbf{320}, 124 (2001)
%
\bibitem{bekki03}
Bekki, K., Harris, W.~E., Harris, G. L. H., et al.: MNRAS \textbf{338}, 587 (2003)
%
\bibitem{aller07}
Aller, M.~C., Richstone, D.~O.: ApJ \textbf{665}, 120 (2007)
%
\bibitem{sarzi01}
Sarzi, M., Rix., H.-W., Shields, J.~C., et al.: ApJ \textbf{550}, 65 (2001)
%
\bibitem{hitschfeld08}
Hitschfeld, M., Aravena, M., Kramer, C., et al.: A\&A \textbf{479}, 75 (2008)
%
\bibitem{riffeser08}
Riffeser, A., Seitz, S., Bender, R.: ApJ \textbf{684}, 1093 (2008)
%
\bibitem{israel09}
Israel, F.~P.: A\&A \textbf{493}, 525 (2009)

\bibitem{atkinson05}
Atkinson, J.~H., Kazeminejad, B., Gaborit, V., et al.: MNRAS \textbf{359}, 504
%
\bibitem{sofue98}
Sofue, Y.: PASJ \textbf{50}, 227 (1998)
%
\bibitem{koda02}
Koda, J., Sofue, Y., Kohno, K., et al.: ApJ \textbf{573}, 105 (2002)
%
\bibitem{wandel99}
Wandel, A.: ApJ \textbf{519}, L39 (1999)
%
\bibitem{genzel95}
Genzel, R., Weitzel, L., Tacconi-Garman, L.~E., et al.: ApJ \textbf{444}, 129 (1995)
%
\bibitem{wold06}
Wold, M., Lacy, M., K\"{a}ufl, H.~U., et al.: A\&A \textbf{460}, 449 (2006)
%
\bibitem{freeman07}
Freeman, K.~C.: Proceedings of IAU Sympsium \#245, edited by Bureau, M., Athanassoula, E, \& Barbuy, B., Univ. Press, Cambridge, p.3 (2007)
%
\bibitem{bosch97}
van den Bosch, F.~C., Jaffe, W.: ASP Conference Series, Vol. 116, edited by Arnaboldi, M., Da Costa, G. S., \& Saha, P., p.142 (1997)
%
\bibitem{killen86}
Killeen N.~E.~B., Bicknell, G.~V., Carter, D.: ApJ \textbf{309}, 45 (1986)
%
\bibitem{rickes09}
Rickes, M.~G., Pastoriza, M.~G., Bonatto, C.: A\&A \textbf{505}, 73 (2009)
%
\bibitem{gultekin09a}
Gultekin, K., Richstone, D.~O., Gebhardt, K., et al.: ApJ \textbf{695}, 1577 (2009)
%
\bibitem{schinnerer02}
Schinnerer, E., Maciejewski, W., Scoville, N., Moustakas, L.~A.: ApJ \textbf{575}, 826 (2002)
%
\bibitem{hu08}
Hu, J.: MNRAS \textbf{386}, 2242 (2008)
%
\bibitem{vandenbergh09a}
van den Bergh, S.: ApJL \textbf{694}, L120 (2009)
%
\bibitem{vandenbergh09b}
van den Bergh, S.: ApJ \textbf{702}, 1502 (2009)
%
\bibitem{moore96}
Moore, B., Katz, N., Lake, G., et al.: Nature \textbf{379}, 613 (1996)
%
\bibitem{curir93}
Curir, A., de Felice, F., Busarello, G., Longo, G., ApL\&C \textbf{28}, 323 (1993)
%
\bibitem{delucia07}
De Lucia, G., Blaizot, J.: MNRAS \textbf{375}, 2 (2007)
%
\bibitem{bower06}
Bower, R.~G., Benson, A.~J., Malbon R., et al.:  MNRAS \textbf{370}, 645 (2006)
%
\bibitem{springel05}
Springel, V., White, S.~D.~M., Jenkins, A., et al.: Nature \textbf{435}, 629 (2005)
%
\bibitem{spergel03}
Spergel, D.~N., Verde, L., Peiris, H.~V., et al.: ApJS \textbf{148}, 175 (2003)
%
\bibitem{lemson06}
Lemson, G., Virgo Consortium t.: preprint, astro-ph/060801 (2006)
%
\bibitem{croton06}
Croton, D.~J., Springel, V., White, S.~D.~M. et al.: MNRAS \textbf{365}, 11 (2006)
%
\bibitem{bonoli09}
Bonoli, S., Marulli, F., Springel, V., et al.: MNRAS \textbf{396}, 423 (2009)
%
\bibitem{marulli09}
Marulli, F., Bonoli, S., Branchini, E., et al.: MNRAS \textbf{396}, 1404 (2009)
%
\bibitem{malbon07}
Malbon, R.~K., Baugh, C.~M., Frenk, C.~S., et al.: MNRAS \textbf{382}, 1394 (2007)
%
\bibitem{baes03}
Baes, M., Buyle, P., Hau, G.~K.~T., et al.: MNRAS \textbf{341}, L44 (2003)
%


\end{thebibliography}

\end{document}